\documentclass{epl}

\title{Instability Criteria for Vibrations of a Vortex Subject to a Gyro-force}
\shorttitle{Instability of a Vortex Vibrations}
\author{B. A. Ivanov\inst{1} \and A. Yu. Galkin\inst{2}}
\institute{
  \inst{1} Institute of Magnetism, National Academy of Sciences and
Ministry of Education of Ukraine, 36(b) Vernadskii avenue, 03142
Kiev, Ukraine \\
  \inst{2} Institute of Metal Physics, National Academy of
Sciences of Ukraine, 36 Vernadskii avenue, 03142 Kiev, Ukraine }


\begin{document}

\maketitle

\begin{abstract}
Here we study the instability of the vortex motion caused by
vortex velocity non-monotonous dependence of the friction force in
the presence of a gyro-force. We demonstrate that even a weak
gyro-force renders the condition for onset of instability more
strict than for a vortex not subject to gyro-forces: either the
friction force must exceed a threshold, or its non-monotonic
velocity dependence must be a rather steep function.
\end{abstract}

The linear topological defects, vortices are one of the most
widely studied objects in modern physics.  The vortex instances
can be found in conventional \cite{DeGennes} and high-temperature
\cite{HTcS} superconductors, superfluid liquids \cite{Donnely},
dilute Bose-Einstein condensates \cite{BEC1,BEC2, Zhou}, Bloch
lines in the domain walls of ferromagnets
\cite{Malozemoff,Baryakhtar}, magnetic vortices in ferromagnets
and antiferromagnets \cite{BarIvKhalat,KosIvKovPR}, and even
singular lines in non-linear optical beams (optical vortices)
\cite{Kivshar}.  For the vortex dynamics description the classical
concept of the elastic string is widely employed. The vortex
possess elastic properties of the string and many peculiarities of
the string dynamics are manifested in the vortex dynamics.  It is
well-known that if the friction force acting on the string depends
non-monotonically on the string velocity, instability of the
string motion results \cite{LordR,Andronov}. When the string moves
in a medium where the friction force acting on it, $\vec F =
-(\vec V /V)F(V) $, depends non-monotonically on velocity within
some range, and moves with a velocity $V_0$ such that
$dF(V)/dV|_{V_0}<0$, then small amplitude vibrations of the
string, superimposed on this background motion, can be described
as linear  vibrations under negative friction. The elastic string
provides one of the best-known examples of auto-oscillation. The
auto-oscillations of a violin string, when the violin string is
under action of a bow moving with a velocity $V_0$, are an
example.

For vortices in condensed media a non-monotonous dependence $F(V)$
is often realized. The physical mechanisms can be very various:
e.g. quasi-particle spectra rearrangement in the core of a moving
Abrikosov vortex in superconductors \cite{LarkinOvch}, vortex
interactions with a system of microdefects \cite{GalIv} or
Cherenkov radiation of sound \cite{Baryakhtar,BarIvKhalat}. For
these cases a non-linear friction force, additional to the
conventional viscous force, with the vortex velocity dependence
$1/V^n$, where $ n $ varies from 0.5 \cite{GalIv} to 2
\cite{LarkinOvch,Baryakhtar,BarIvKhalat} appears.

An important feature common to dynamics of many types of vortices
is the presence of a gyro-force, see references \cite{DeGennes}
-\cite{Kivshar}. The gyro-force is formally equivalent to the
Lorentz force that acts on a charged  elastic string  placed in
the magnetic field parallel to the string line.

  A gyro-force $|\vec F_g | =
G\cdot V$ ($G$ is the gyro-constant) directed perpendicular to the
vortex velocity $\vec V$ is an inherent feature of vortex
dynamics. Experimentally it has been observed in ferromagnets
\cite{Malozemoff,Baryakhtar}, in high-Tc superconductors in
super-clean limit \cite{Harris}, and in optics \cite{Kivshar}. The
presence of a gyro-force for these media is a direct consequence
of the topological character of the vortex
\cite{Malozemoff,Baryakhtar, BarIvKhalat}.  For models having
Lorentz-invariant dynamics a gyro-force is not an inherent feature
of the vortex. An example of such a model is a nonlinear
sigma--model, used for description of antiferromagnets
\cite{Baryakhtar, BarIvKhalat} and quantum nematic phases in
Bose-Einstein condensates \cite{Zhou} and spin--1 non-Heisenberg
magnets \cite{IvKol02}. Even for these cases, a gyro-force could
appear in the presence of an external magnetic field parallel to
the vortex line \cite{Ivanov94, IvKol95}.

In this Letter we demonstrate that the question of the general
stability of the motion of string-like objects (vortices) in
viscous media needs to be revised in the presence of a gyro-force.
It is shown that in this case the presence of non-monotonic
dependence of the friction force, as well as the negative
differential mobility does not serve as a stability condition and
more complicated stability conditions arise. We will use the
simple model of the elastic string, which has a gyro-force, which
is a rather general one and its applicability falls outside the
framework of the theory of vortices in ordered media.

Consider the vortex directed along the z-axis. Its dynamics may be
described on the basis of the effective equation for a
two-dimensional vector, $\vec U = \vec U (z,t)$ lying in the
$(x,y)$ plane and determining the displacement of the vortex
center:
\begin{equation}
\label{MotionEq}
 m \ddot{\vec{U}}-\sigma (\partial^2 \vec
{U}/\partial z^2)+G(\widehat{\vec {z}}\times
\dot{\vec{U}})=\vec{F}.
\end{equation}

Here $\dot{\vec{U}}\equiv \partial{\vec{U}}/\partial{t}$, $m$, $G$
and $\sigma$ are the vortex mass, the gyro-force constant and the
energy per unit length of the vortex, respectively. The friction
force $\vec{F}\equiv
\vec{F}(\dot{\vec{U}})=-(\dot{\vec{U}}/|\dot{\vec{U}}|)F(|\dot{\vec{U}}|)
$ is considered to be collinear to the instantaneous velocity of
the vortex with positive value of $F(V)$.

In order to investigate the stability of the motion of the vortex
when moving with a velocity close to $V_0$ we assume $ \vec U =
\vec V_0 t + \vec u (z,t)$, where $\vec u(z,t)$ is a small
perturbation of the steady motion at $\vec V_0$, and linearize
(\ref{MotionEq}) (in effect, the friction force $\vec F$) over
$\vec u$. In this linear approximation the friction force can be
represented as a sum of components perpendicular and parallel to
$\vec V_0$:
\begin{equation}
 \vec F= -\vec V_0(\dot {\vec u} \vec V_0)(F^{\prime } /
V_0^2)-(\vec V_0 \times (\dot {\vec u} \times \vec V_0))(F /
V_0^3), \nonumber
\end{equation}
where $F=F(V_0),F^{\prime }=[dF(V)/dV]_{V=V_0}$. Then we can find
the solution in Fourier representation, $\vec u(z,t)\propto \vec
u_q \exp(\Lambda t +iqz)$ , where $\Lambda $ defines the character
of the time evolution of small deviations from the steady motion.
If the real part of $\Lambda$ is negative then the value
$-$Re$\Lambda>0$ serves as a damping coefficient for small
vibrations of the vortex. Otherwise, if the real part of $\Lambda$
is positive for some values of the wave number $q$, weak
vibrations cause instability with the incremental factor
Re$\Lambda>0$.

If the gyro-force is absent the situation is very simple. The
equations for the components of $\vec u$ parallel and
perpendicular to the $\vec V_0$ are then independent. They have
different effective friction coefficients $F^{\prime }$ and $F/V$.
The value of $F/V$ is necessarily positive while $F^{\prime }$ can
be either positive or negative within any given small velocity
range. It is therefore evident that for $G = 0$ vibrations with
$\vec u \bot \vec V_0$ are damped, while the possible instability
of vibrations with $\vec u
\parallel \vec V_0$ is related to the condition $F^{\prime } < 0$.
Thus for $G = 0$ the sole criterion for instability of steady
motion of the vortex at velocity $\vec V_0$ is
$[dF(V)/dV]_{V=V_0}<0$. If $G\neq 0$, then oscillations with $\vec
u \bot \vec V_0$ and $\vec u \parallel \vec V_0$ become coupled
and the characteristic equation for $\Lambda$ is then a 4-th power
equation. It can be presented in the simple form:
\begin{equation}
\nonumber [m \Lambda^2 + (\Lambda/2)(F^{\prime }+F/V)+\sigma
q^2]^2 =    (\Lambda/2)^2[(F^{\prime } -F/V)^2-4G^2],
\end{equation}
and solved exactly. The analysis demonstrates that the real parts
of all four roots of this equation are negative, i.e. the motion
is stable when two criteria are \emph{simultaneously} satisfied:
${F}^{\prime }(F/V) + G^2
> 0$ and ${F}^{\prime }+ (F/V) > 0$. (It is evident that these criteria do
not depend on the mass and elasticity.) Thus, two
\emph{independent} criteria for instability exist, which
correspond to violation of one or other of these two stability
criteria.  The first condition for the vortex motion to be
unstable can be written as
\begin{equation}
\label{Cond1}
    -( {F}^{\prime } )(F/V) > G^2.
\end{equation}
In the limit $G \rightarrow 0$ this condition coincides with the
inequality ${F}^{\prime }<0$.  The second instability condition
\begin{equation}
\label{Cond2}  F^{\prime }+ (F/V) < 0
\end{equation}
can be satisfied however weak the friction and however large the
gyro-constant $G$. Therefore this second criterion can also be
significant for the case of interest when $G$ is non-small.

Suppose that the friction force $F(V)$ is the sum of a regular
viscous friction force with a constant velocity coefficient $\beta
$ and a non-linear friction force $f(V)$, which can be a
decreasing function of $V$, so that $F(V) =  \beta V + f(V)$.  The
condition $ F^{\prime }<0 $ requires the value of $f^{\prime }
(V)$ to be negative and that the maximal value of its modulus,
reached at $V \cong V_m$, exceeds $\beta$, $|f^{\prime
}|^{max}>\beta$. It is evident that the condition $ F^{\prime } <0
$ can be met however weak the friction coefficient $\beta \sim
|f^{\prime }|^{max}\approx (f/V)\ll G$; it is only important that
the values $\beta$ and $|f^{\prime }|^{max} \sim f/V$ should be
comparable. However, at fairly large $G$, instability condition
(\ref{Cond1}) is certainly more restrictive as it involves $G$.
Estimating $|f^{\prime }|^{max}\sim f(V_m)/V_m $, one can express
the first instability condition (\ref{Cond1}) as $|f^{\prime
}|^{max} > \sqrt{G^2+\beta^2}$, which cannot be fulfilled for
large enough $G$ and however weak the friction force $f$.

The instability condition (\ref{Cond1}) can acquire a clear
physical meaning if one introduces the non-linear differential
mobility $\mu (V)$ in respect to the external force $F_e$,
$\mu(V)=dV(F_e)/dF_e $. In fact, the motion under action of a
constant external force takes place at the Hall angle $ \alpha_H$
regarding to $\vec F_e $, and the velocity is determined by the
condition $G^2V^2+F^2(V)=F^2_e$. The $\mu(V) $ can be presented as
\begin{equation}
\nonumber \mu(V)= \frac{\sqrt{F^2(V)+G^2V^2}}{VG^2+F(V)F^{\prime
}(V)}
\end{equation}
and the instability condition (\ref{Cond1}) can be considered as
the negative differential mobility $\mu(V) < 0$ . The condition of
the negative differential mobility for small friction and finite
$G$ could be met with difficulty, even if $ F^{\prime } < 0$.

Thus, for this case - large $G$, the condition (\ref{Cond2})
provides the better opportunity for instability: it does not
involve $G$ and can be met at low friction. However, there is
another strong restriction that can prevent instability. If one
extrapolates the function $f(V)$ by a power function, $ f(V)= \eta
/V^n$, where $\eta $ is some coefficient, then $F^{\prime }$  can
be negative at any ratio between $\eta $ and $\beta$ and at any $n
> 0$. However, the condition (\ref{Cond2}) yields
$(1-n)\eta/V^{n+1}+2\beta < 0 $ and is not fulfilled with $n \leq
1$. So here at any small $G$ instability requires a rather steep
dependence of $f(V)$.

As an example, for the dynamics of a single vortex in a media with
weak inhomogeneities, $n =1/2$  \cite{GalIv}, the value of
$f^{\prime }+f/V=f/2V$ is always positive and the stability of the
motion is not destroyed by the condition (\ref{Cond2}).
Nevertheless, when a dense vortex lattice moves through defects
 $n=2$ and instability due to the criterion
(\ref{Cond2}) could appear \cite{GalkIvan}.

In summary, the onset of instability of motion of a vortex subject
to a gyro-force cannot be attributed merely neither to negative
incremental friction $F^{\prime }<0 $ nor of the negative
differential mobility. It requires additional restrictions on the
velocity dependence of the friction force $F(V)$ or on the value
of the gyro-force $G$. In fact, two independent instability
criteria arise. One of them is associated with appearance of the
negative differential mobility, $dV/dF_e$ and can be discussed as
stability with respect to the \emph{longitudinal} perturbations.
The other condition is nothing to do with differential mobility
and corresponds to the negative effective dumping coefficient
averaged over the longitudinal and \emph{transverse} (regarding to
the velocity $\vec V$) oscillations of the vortex. For $G = 0$ the
condition $dV/dF_e < 0 $ coincides with $F^{\prime }<0 $, however
at $F^{\prime },F(V)/V \ll G $ these two conditions are
essentially different.

It is worth to note that most of above considered examples are
related to vortex-like linear defects, especially vortices in
superconductors and magnets. Nevertheless, the model of the
string, which has a gyro-force is a rather general one and falls
outside the framework of the theory of vortices in ordered media.

\acknowledgments We thank J. E. L. Bishop and C.E.Zaspel for their
useful comments on the manuscript. One of us (B.I.) thanks the
grant I/75895 from the Volkswagen-Stiftung  for partial support.

\end{document}